\documentclass[pra,twocolumn,superscriptaddress,showpacs,amsmath,amstex,amssymb,citeautoscript]{revtex4-1}

\usepackage[T1]{fontenc}
\usepackage[utf8]{inputenc}
\usepackage{lipsum}
\usepackage{amsmath}
\usepackage{amssymb}
\usepackage{bbm}
\usepackage{braket}
\usepackage{xcolor}
\usepackage{pifont}
\usepackage[mathscr]{euscript}
\usepackage[shortlabels]{enumitem}
\usepackage[justification=justified]{subcaption}
\usepackage{graphicx}
\usepackage{lipsum}
\allowdisplaybreaks
\usepackage{float}
\usepackage{graphicx}
\usepackage{dsfont}
\usepackage{comment}
\usepackage[colorlinks=true]{hyperref}  
\hypersetup{
    bookmarks=true,         
    unicode=false,          
    pdftoolbar=true,        
    pdfmenubar=true,        
    pdffitwindow=false,     
    pdfstartview={FitH},    
    pdftitle={Enhanced Superconducting Diode Effect},    
    pdfauthor={Banerjee and Scheurer},     
    pdfsubject={},   
    pdfcreator={},   
    pdfproducer={}, 
    pdfkeywords={} {} {}, 
    pdfnewwindow=true,      
    colorlinks=true,       
    linkcolor=blue, 
    citecolor=blue,        
    filecolor=magenta,      
    urlcolor=blue           
}

\newcommand{\equref}[1]{Eq.~(\ref{#1})}
\newcommand{\equsref}[2]{Eqs.~(\ref{#1}) and (\ref{#2})}

\newcommand{\figref}[1]{Fig.~\ref{#1}}
\newcommand{\refcite}[1]{Ref.~\onlinecite{#1}}

\newcommand{\appref}[1]{Appendix~\ref{#1}}

\newcommand{\pdagger}{{\phantom{\dagger}}}
\newcommand{\diff}{\mathrm{d}}

\renewcommand{\approx}{\simeq}

\renewcommand{\vec}[1]{\boldsymbol{#1}}

\definecolor{wrongultramarine}{rgb}{1,0.5,0}

\begin{document}

\title{Enhanced Superconducting Diode Effect due to coexisting Phases}

\author{Sayan Banerjee}
\affiliation{Institute for Theoretical Physics, University of Innsbruck, Innsbruck A-6020, Austria}

\author{Mathias S.~Scheurer}
\affiliation{Institute for Theoretical Physics, University of Innsbruck, Innsbruck A-6020, Austria}

\begin{abstract}
The superconducting diode effect refers to an asymmetry in the critical supercurrent $J_c(\hat{n})$ along opposite directions, $J_c(\hat{n})\neq  J_c(-\hat{n})$. While the basic symmetry requirements for this effect are known, it is, for junction-free systems, difficult to capture within current theoretical models the large current asymmetries $J_c(\hat{n})/J_c(-\hat{n})$ recently observed in experiment. We here propose and develop a theory for an enhancement mechanism of the diode effect arising from spontaneous symmetry breaking. We show---both within a phenomenological and a microscopic theory---that there is a coupling of the supercurrent and the underlying symmetry-breaking order parameter. This coupling can enhance the current asymmetry significantly. Our work might not only provide a possible explanation for recent experiments on trilayer graphene but also pave the way for future realizations of the superconducting diode effect with large current asymmetries.
\end{abstract}

\maketitle
Diodes, which are characterized by a highly asymmetric relation between resistance $R$ and current $\vec{J}$, $R(\vec{J}) \neq R(-\vec{J})$, are an integral part of modern-day electronics. A superconductor where the critical current $J_c$ is different along opposite directions, $J_c(\hat{n}) > J_c(-\hat{n})$, can realize a superconducting analogue of a diode in the sense that $R(-J\hat{n}) \neq R(J\hat{n})=0$, if $J_c(-\hat{n}) < J <J_c(\hat{n})$. While asymmetries in superconducting current-voltage relations in low-symmetry superconductors had been observed before, see, e.g., \cite{broussard_critical_1988,jiang_asymmetric_1994,papon_asymmetric_2008}, the potential technological applications of and fundamental scientific questions associated with this superconducting diode effect (SDE) have attracted significant experimental \cite{ando_observation_2020,lyu_superconducting_2021,du_superconducting_2023,sundaresh_diamagnetic_2023,kealhofer_anomalous_2023,hou_ubiquitous_2023,chen_superconducting_2023,gupta_superconducting_2022,AnotherJosephson,banerjee_phase_2023,pal_josephson_2022,kim_intrinsic_2023,gupta_superconducting_2022,MoreJosephson,ParadisoNbSe2,wu_field-free_2022,diez-merida_magnetic_2021,golod_demonstration_2022,chiles_non-reciprocal_2022-1,zhang_reconfigurable_2023,shin_magnetic_2021,jeon_zero-field_2022,kealhofer_anomalous_2023,hou_ubiquitous_2023,lin_zero-field_2022,anwar_spontaneous_2022,narita_field-free_2022,gutfreund_direct_2023} and theoretical \cite{daido_superconducting_2022,daido_intrinsic_2022,yuan_supercurrent_2022,he_phenomenological_2022,ilic_theory_2022,scammell_theory_2022,zinkl_symmetry_2022,PhysRevX.12.041013,he_supercurrent_2022,PhysRevB.106.L140505,jiang_field-free_2022,kokkeler_field-free_2022,chazono_piezoelectric_2022,vodolazov_superconducting_2005,de_picoli_superconducting_2023,kochan_phenomenological_2023,ikeda_intrinsic_2022,tanaka_theory_2022,wang_symmetry_2022,haenel_superconducting_2022,legg_parity_2023,cuozzo_microwave-tunable_2023,souto_josephson_2022,cheng_josephson_2023,steiner_diode_2022,costa_microscopic_2023,wei_supercurrent_2022,legg_superconducting_2022,karabassov_hybrid_2022,2022arXiv221114846H} attention in recent years; this led to a variety of different realizations both in tunnel-junction setups \cite{chen_superconducting_2023,gupta_superconducting_2022,AnotherJosephson,banerjee_phase_2023,pal_josephson_2022,kim_intrinsic_2023,gupta_superconducting_2022,MoreJosephson,ParadisoNbSe2,wu_field-free_2022,diez-merida_magnetic_2021,golod_demonstration_2022,chiles_non-reciprocal_2022-1,zhang_reconfigurable_2023,shin_magnetic_2021,jeon_zero-field_2022} and in single junction-free superconducting phases \cite{ando_observation_2020,lyu_superconducting_2021,du_superconducting_2023,sundaresh_diamagnetic_2023,kealhofer_anomalous_2023,hou_ubiquitous_2023,lin_zero-field_2022,anwar_spontaneous_2022,narita_field-free_2022,gutfreund_direct_2023}. Importantly, critical current asymmetries in superconductors require broken time-reversal symmetry (TRS), which can not only be achieved by applying a magnetic field \cite{chen_superconducting_2023,gupta_superconducting_2022,AnotherJosephson,banerjee_phase_2023,pal_josephson_2022,kim_intrinsic_2023,gupta_superconducting_2022,MoreJosephson,ParadisoNbSe2,ando_observation_2020,lyu_superconducting_2021,du_superconducting_2023,sundaresh_diamagnetic_2023,kealhofer_anomalous_2023,hou_ubiquitous_2023}, a current \cite{chiles_non-reciprocal_2022-1,zhang_reconfigurable_2023}, or magnetic proximity \cite{shin_magnetic_2021,narita_field-free_2022,gutfreund_direct_2023,jeon_zero-field_2022}, but also result from \textit{spontaneous} TRS-breaking in a junction \cite{wu_field-free_2022,diez-merida_magnetic_2021,golod_demonstration_2022} or homogeneous superconductor \cite{lin_zero-field_2022,anwar_spontaneous_2022}; aside from the fundamental interest in the competition of magnetism and superconductivity in a single electron liquid, this might also be useful for integrated designs where external fields are not practical. Another crucial aspect for applications is the degree of current asymmetry, conveniently measured by the dimensionless diode efficiency $\eta(\hat{n}) = |J_c(\hat{n})- J_c(-\hat{n})|/( J_c(\hat{n})+ J_c(-\hat{n}))$. A large efficiency, as close to $1$ as possible, is desirable for the system to operate as a superconducting diode without having to fine-tune the current magnitude. 

In this work, we propose and demonstrate theoretically an enhancement mechanism for the SDE efficiency $\eta$, specifically for systems with spontaneously broken TRS. While it is clear by symmetry that the condensation of a time-reversal-odd order parameter can affect the critical supercurrent and induce a SDE, we here demonstrate that the supercurrent also couples back to the underlying symmetry-breaking order parameter. If superconductivity and the TRS-breaking order have similar energy scales, this ``back action’’ can be quite large and, as we demonstrate, can enhance $\eta$ significantly. Although this mechanism is more generally valid and should apply to different models and forms of the time-reversal-odd order parameter, we use a model inspired by graphene moiré systems; this is motivated by \refcite{lin_zero-field_2022} where a junction-free sample of twisted trilayer graphene was studied and a zero-field SDE with a particularly large $\eta$ was observed. Our work, thus, both provides a possible route to explaining the large $\eta$ of \refcite{lin_zero-field_2022} and paves the way for the design of zero-field superconducting diodes with large current asymmetries.

\vspace{1em}
\textit{Model for SDE.---}As motivated above, we study an electronic Hamiltonian of the form
\begin{align}\begin{split}
    \mathcal{H}_c = &\sum_{\vec{k},\eta} \xi_{\vec{\vec{k}},\eta} c^\dagger_{\vec{k},\eta}c^\pdagger_{\vec{k},\eta} +  \Phi_V \sum_{\vec{k},\eta} \eta c^\dagger_{\vec{k},\eta}c^\pdagger_{\vec{k},\eta} \\ &+ \sum_{\vec{k},\vec{q}} \left[ \Delta_{\vec{q}} c^\dagger_{\vec{k}+\vec{q}/2,+}c^\dagger_{-\vec{k}+\vec{q}/2,-}  + \text{H.c.} \right], \label{CouplingToVPandSC}
\end{split}\end{align}
where $c_{\vec{k},\eta}^\dagger$ creates an electron in valley $\eta=\pm$ and at momentum $\vec{k}$. The three terms in $\mathcal{H}_c$ capture the non-interacting bandstructure of the nearly flat bands at the chemical potential, which obey $\xi_{\vec{\vec{k}},\eta} = \xi_{-\vec{\vec{k}},-\eta} \equiv \xi_{\vec{k}\cdot \eta}$ due to TRS, the coupling of the TRS-breaking normal state order (in our case, valley imbalance $\Phi_V$) and of the superconducting order parameter to the electrons, respectively. For notational simplicity, we suppress the spin index in \equref{CouplingToVPandSC} and in the following, but emphasize that our results apply both for spin polarized bands \cite{morissette_electron_2022,zondiner_cascade_2020} as well as in the spinful case, with singlet or triplet pairing.

To stabilize superconductivity, we take an attractive ($g_c > 0$) interaction $\mathcal{H}_{g_{c}} = -\frac{g_c}{2} \sum_{\vec{q},\eta,\eta'} C^\dagger_{\vec{q};\eta,\eta'} C^\pdagger_{\vec{q};\eta,\eta'}$ with $C_{\vec{q};\eta,\eta'} = \sum_{\vec{k}} c_{\vec{k}+\vec{q}/2,\eta} c_{-\vec{k}+\vec{q}/2,\eta'}$ and perform a decoupling in the intervalley pairing channel, $\braket{C_{\vec{q};\eta,\eta'}} = \eta \delta_{\eta,\eta'} \Delta_{\vec{q}}/g_c$; we focus on intervalley pairing not only because it is favored as a consequence of TRS for a finite range of $\Phi_V$ \cite{scammell_theory_2022} but also because intravalley pairing is more susceptible to impurities and, hence, likely less relevant experimentally. The resulting mean-field Hamiltonian for superconductivity for given valley polarization $\Phi_V$ then reads  as $\mathcal{H}_{\text{S}} = \mathcal{H}_c + \sum_{\vec{q}} |\Delta_{\vec{q}}|^2/g_c$. While we will compute the supercurrent systematically below, we start for illustration purposes with a semi-phenomenological approach based on a Ginzburg-Landau expansion. Integrating out the electrons in $\mathcal{H}_{\text{S}}$ and expanding up to second order in $\Delta_{\vec{q}}$, the change $\delta \mathcal{F}_{\text{S}} := \mathcal{F}[\Delta_{\vec{q}},\Phi_V] - \mathcal{F}[0,\Phi_V]$ of the free-energy with superconductivity reads as
\begin{equation*}
    \delta \mathcal{F}_{\text{S}} \sim \sum_{\vec{q}} a^{\text{S}}_{\vec{q}} \, |\Delta_{\vec{q}}|^2 + b^{\text{S}} \sum_{\vec{q}_i} \Delta^*_{\vec{q}_1}\Delta^*_{\vec{q}_2}\Delta^{\phantom{*}}_{\vec{q}_3}\Delta^{\phantom{*}}_{\vec{q}_4} \delta_{\vec{q}_1+\vec{q}_2,\vec{q}_3+\vec{q}_4},
\end{equation*}
where we neglected the momentum dependence of the quartic term. It holds $a^{\text{S}}_{\vec{q}} = 1/g_c - \Gamma_{\vec{q}}$ with \cite{scammell_theory_2022}
\begin{equation}
    \Gamma_{\vec{q}} = \frac{1}{2N}\sum_{\vec{k}} \frac{\tanh{\frac{E_{\vec{k},\vec{q},+}}{2 T}} + \tanh{\frac{E_{\vec{k},\vec{q},-}}{2 T}}}{E_{\vec{k},\vec{q},+} + E_{\vec{k},\vec{q},-}}, \label{ParticleParticleBubble}
\end{equation}
where $N$ is the number of unit cells and $E_{\vec{k},\vec{q},\eta} =  \xi_{\vec{k} + \eta \vec{q}/2} + \eta \,\Phi_V$ encodes the normal-state dispersion, associated with the first line of \equref{CouplingToVPandSC}. 

The equilibrium superconducting state at given $\Phi_V$ is found by minimizing $\delta \mathcal{F}_{\text{S}}$. Restricting the analysis to single-$\vec{q}$ states, $\Delta_{\vec{q}} \propto \delta_{\vec{q},\vec{q}_0}$, the value of $\vec{q}_0$ is determined by the minimum of $a^{\text{S}}_{\vec{q}}$ (the maximum of $\Gamma_{\vec{q}}$). As expected, the current \cite{daido_intrinsic_2022,yuan_supercurrent_2022,scammell_theory_2022}
\begin{equation}
    \vec{J}(\vec{q}) = 2 e \Delta_{\vec{q}} \vec{v}_{\vec{q}}, \quad \vec{v}_{\vec{q}} = \vec{\nabla}_{\vec{q}} a^{\text{S}}_{\vec{q}}, \label{PhenExpressionForTheSupercurrent}
\end{equation}
vanishes in equilibrium, $\vec{J}(\vec{q}_0) = 0$; a supercurrent-carrying state, therefore, corresponds to pairing with $\vec{q}\neq \vec{q}_0$. We define the critical current $J_c(\hat{n})$ along $\hat{n}$ as the maximal magnitude of $\vec{J}(\vec{q})$ oriented along $\hat{n}$. 

For $\Phi_V = 0$, it holds $E_{\vec{k},\vec{q},\eta} = E_{\vec{k},-\vec{q},-\eta}$ and thus $\Gamma_{\vec{q}} = \Gamma_{-\vec{q}}$ in \equref{ParticleParticleBubble}. This, in turn, immediately implies $\vec{J}(\vec{q}) = -\vec{J}(-\vec{q})$ and $J_c(\hat{n}) = J_c(-\hat{n})$, i.e., there is no SDE. This was expected as a consequence of TRS or two-fold rotational symmetry, $C_{2z}$, for that matter---both have to be broken to get $J_c(\hat{n}) \neq J_c(-\hat{n})$. This is achieved by $\Phi_V\neq 0$ \cite{scammell_theory_2022}, leading to a finite SDE efficiency, 
\begin{equation}
    \eta = \max_{\hat{n}} \frac{|J_c(\hat{n})-J_c(-\hat{n})|}{J_c(\hat{n}) + J_c(-\hat{n})} \in [0,1].
    \label{efficiency}
\end{equation}
Although generically non-zero by symmetry, it does not guarantee that $\eta$ can reach high values corresponding to large current asymmetries, $J_c(\hat{n}) \gg J_c(-\hat{n})$---as mentioned above, a property desirable for applications of the SDE and relevant fundamentally to understand recent experiments \cite{lin_zero-field_2022}.

\vspace{1em}
\textit{Phenomenological coupling.---}The main result of this work is that the diode-effect efficiency $\eta$ can be significantly enhanced by taking into account that the strength of the valley polarization is affected by the supercurrent or, put differently, $\Phi_V$ entering $E_{\vec{k},\vec{q},\eta}$ in \equref{ParticleParticleBubble} also depends on the center of mass momentum $\vec{q}$ of the Cooper pairs, $\Phi_V\rightarrow \Phi_V(\vec{q})$. Postponing a systematic microscopic computation that treats $\Delta_{\vec{q}}$ and $\Phi_V(\vec{q})$ on equal footing, we start with a simpler phenomenological analysis. In this simplified description, we determine $\Phi_V(\vec{q})$ by minimizing the free energy
\begin{equation}
     \mathcal{F}_{\text{V}} \sim a_{\Phi}\Phi_{V}^{2} + b_{\Phi} \Phi_{V}^{4} + \Phi_V \sum_{\vec{q}} |\Delta_{\vec{q}}|^2 \left[ F(\vec{q})  + \alpha'' \Phi_V \right]
\label{freenergyexpansion}
\end{equation}
with respect to $\Phi_V$, where terms involving fifth and higher powers of the order parameters $\Phi_V$, $\Delta_{\vec{q}}$ are neglected. 
The first two terms in $\mathcal{F}_{\text{V}}$ determine the value of $\Phi_{V}$ in the absence of superconductivity. Meanwhile, the remaining terms in \equref{freenergyexpansion} capture the key coupling between the momentum of the Cooper pairs and valley polarization. The last term, $\propto \alpha''$, simply describes the fact that increasing (decreasing) the superconducting order parameter magnitude suppresses (enhances) valley polarization. The other term with $F(\vec{q}) = -F(-\vec{q})$ describes the fact that superconductivity at finite $\vec{q}$ breaks TRS and $C_{2z}$ and can, hence, couple to the first (and in general an odd) power of $\Phi_V$. Motivated by twisted graphene systems, which exhibit $C_{3z}$ rotational symmetry, we write
\begin{equation}
    F(\vec{q}) = \alpha \, \hat{\vec{e}}_{\epsilon}\cdot \vec{X}(\vec{q}) + \alpha' \, \chi(\vec{q}), \label{CouplingFunction}
\end{equation}
where $\vec{X}(\vec{q})$ and $\chi(\vec{q})$ are real-valued, periodic on the Brillouin zone, odd in $\vec{q}$, and transform as a vector and scalar under $C_{3z}$, respectively; in the limit of small $\vec{q}$, it holds $\vec{X} \sim (q_x,q_y)^T$ and $\chi(\vec{q}) \sim q_x(q_x^2-3q_y^2)$, using a convention where one of the primitive translation vectors, $\vec{a}_j$, of the underlying triangular Bravais lattice points along the $x$ direction. For the following calculations, we extend these functions to the Brillouin zone using the leading lattice harmonics (see \appref{AppPhenTheory}). Importantly, the first term in \equref{CouplingFunction}, requires finite strain or nematic order which break $C_{3z}$ symmetry explicitly. These phenomena, which seem to be ubiquitous in graphene moiré systems \cite{jiang_charge_2019, kerelsky_maximized_2019,cao_nematicity_2021,rubio-verdu_moire_2022,samajdar_electric-field-tunable_2021, sobral_machine_2023,zhang_electronic_2022-1}, define a preferred in-plane direction, captured by the unit vector $\hat{\vec{e}}_{\epsilon}$. For instance, in the case of strain, $\hat{\vec{e}}_{\epsilon} \propto (\epsilon_{xx}-\epsilon_{yy}, \epsilon_{xy}+\epsilon_{yz})^T$, where $\epsilon_{ij}$ are the strain-tensor elements. We note that the coupling would also be present when $\Phi_V$ is replaced by another order parameter that itself already breaks both TRS and $C_{3z}$ simultaneously \cite{zhang_electronic_2022-1,lin_spontaneous_2023}. 
The second term in \equref{CouplingFunction}, albeit subleading for small $\vec{q}$, is finite regardless of whether $C_{3z}$ is broken.

To study the SDE, we self-consistently minimize $\mathcal{F}_V[\Phi_{V},\Delta_{\vec{q}}]$ and $\delta \mathcal{F}_{\text{S}}[\Phi_{V},\Delta_{\vec{q}}]$ with respect to $\Phi_{V}$ and $\Delta_{\vec{q}}$, respectively, for given $\vec{q}$, and compute the current according to \equref{PhenExpressionForTheSupercurrent}. For concreteness, we use a nearest-neighbor dispersion with finite flux $\phi$ on the triangular lattice, $\xi_{\vec{k}} = - \sum_{j=1}^3 t_j \cos (\vec{a}_j\cdot \vec{k}-\phi/3)$ where $\vec{a}_j$ are three $C_{3z}$-related primitive vectors. We choose $t_1=t$, $t_2=t_3=t(1+\beta)$ to parameterize the impact of strain ($\propto \beta$) on the dispersion and, for notational simplicity, measure all energies in units of $t$ in the following.

\begin{figure}[tb]
   \centering
    \includegraphics[width=\linewidth]{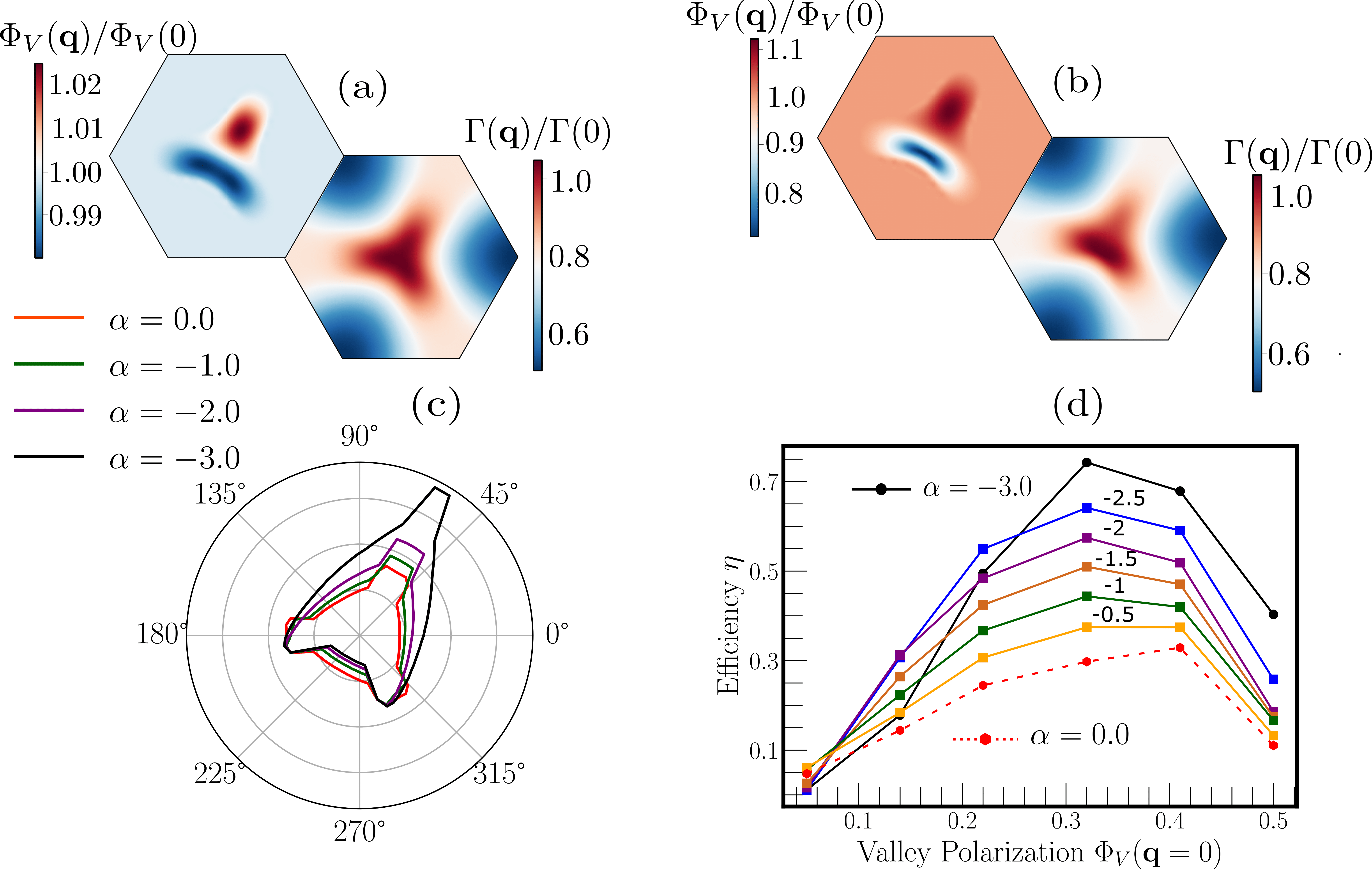}
    \caption{Phenomenological theory with leading nematic coupling, $\alpha \neq 0$, $\alpha'=\alpha''=0$ in \equref{freenergyexpansion}. $\Phi_v(\vec{q})$ and $\Gamma(\vec{q})$ for (a) $\alpha = -0.5$ and (b) $\alpha = -3.0$ are shown in upper and lower panels, respectively. (c) Angular ($\hat{n}$) dependence of $J_c$ for different $\alpha$, taking $|\Phi(\vec{q}=0)|=0.32$. (d) SDE efficiency $\eta$ for different $\alpha$ as a function of valley polarization at $\vec{q}=0$. We use $g_c^{-1}=0.8\Gamma_{\vec{q}=0}[\Phi_V(\vec{q}=0)]$, $\mu = -0.68$, $\phi=-0.7\pi$, $T=0.2$, $\hat{\vec{e}}_{\epsilon} = (\cos{\pi/3},\sin{\pi/3})$.}
    \label{fig:PhenTheoryLinearCoupling}
\end{figure}

\vspace{1em}
\textit{Back action and SDE.---}We start with the first term in \equref{CouplingFunction}, i.e., set $\alpha'=\alpha''=0$ in $\mathcal{F}_V$. As can be seen in \figref{fig:PhenTheoryLinearCoupling}(a), finite $\alpha$ now induces a $\vec{q}$ dependence in $\Phi_V$, which breaks $C_{3z}$ symmetry as a result of the preferred direction associated with $\hat{\vec{e}}_{\epsilon}$. As we set $\beta=0$ for now, $\Gamma_{\vec{q}}$ looks virtually $C_{3z}$ symmetric, while larger $\alpha$ will lead to a noticeable asymmetry in $\Gamma_{\vec{q}}$ as well, see \figref{fig:PhenTheoryLinearCoupling}(b). This results from a ``back action mechanism'', where a finite supercurrent, which is associated with detuning the center of mass momentum $\vec{q}_{0} \rightarrow \vec{q} +\delta \vec{q}_{0}$ of the Cooper pairs away from their equilibrium value $\vec{q}_{0}$, changes the strength of $\Phi_V$; as increasing (decreasing) $\Phi_V$ will weaken (strengthen) superconductivity, this, in turn, influences $\Gamma_{\vec{q}}$. This can be thought of as the supercurrent analogue of the coupling between the dissipative current and valley polarization observed in graphene moiré systems \cite{serlin_intrinsic_2020, sharpe_emergent_2019, ying_current_2021}. We note in passing that a sufficiently strong supercurrent can also flip the sign $\Phi_V$ in our theory.

This back action also affects the $\vec{q}$ dependence of the supercurrent, which allows to enhance the SDE as can be seen in \figref{fig:PhenTheoryLinearCoupling}(c). We further present in \figref{fig:PhenTheoryLinearCoupling}(d) the diode efficiency $\eta$ as a function of $\Phi_{V}(\vec{q}=0)$ for different strengths of the back action. Without back action ($\alpha=0$, red dashed line), $\eta$ does not exceed $30\%$, corresponding to $J_c(\hat{n}) \approx 1.86 J_c(-\hat{n})$. In contrast, increasing $\alpha$ and, thus, boosting the back action mechanism can significantly enhance the SDE, even reaching efficiencies as high as 75\% (solid black line), i.e., $J_c(\hat{n}) \approx 7 J_c(-\hat{n})$.

\begin{figure}[tb]
   \centering
    \includegraphics[width=\linewidth]{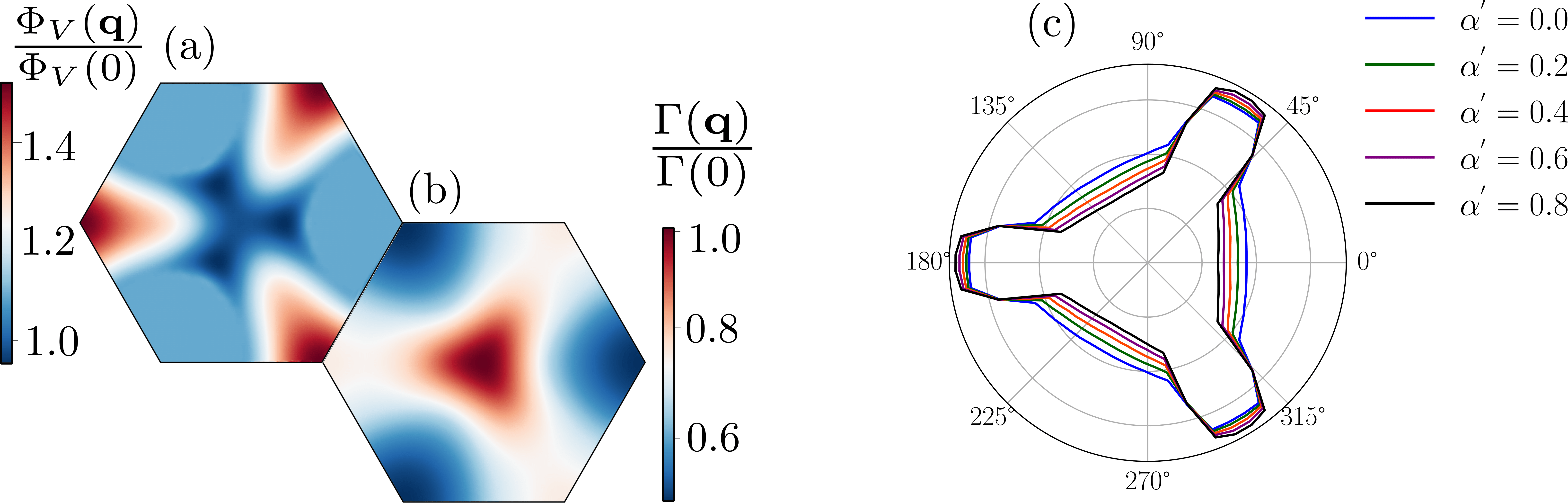}
    \caption{Phenomenological theory with $C_{3z}$-symmetric coupling, $\alpha'\neq 0$, $\alpha=0$, $\alpha'' = 1.0$ in \equref{freenergyexpansion}. $\Phi_v(\vec{q})$ and $\Gamma(\vec{q})$ in (a) and (b) are shown for $\alpha'=0.8$ and (c) displays the angular ($\hat{n}$) dependence of $J_c$ for different $\alpha'$. We use $g_{c}^{-1}=0.7\Gamma_{\vec{q}=0}[\Phi_V=0.32]$, $\mu = -0.68$, $\phi=-0.7\pi$, $T=0.2$.}
    \label{fig:PhenTheoryCubicCoupling}
\end{figure}

This enhancement mechanism of the SDE is also possible in the presence of $C_{3z}$, where $\alpha=0$, if we take into account finite $\alpha'$ in \equref{CouplingFunction}, see \figref{fig:PhenTheoryCubicCoupling}. We can clearly see the induced $\vec{q}$ (and thus supercurrent) dependence of $\Phi_V$ in \figref{fig:PhenTheoryCubicCoupling}(a). However, the back action onto $\Gamma_{\vec{q}}$ is less clearly seen in \figref{fig:PhenTheoryCubicCoupling}(b) at a single $\alpha'$, since it is $C_{3z}$ symmetric with and without it. Most importantly, though, \figref{fig:PhenTheoryCubicCoupling}(c) reveals that also the subleading back action ($\alpha=0$) enhances the current asymmetry.

\vspace{1em}
\textit{Self-consistent theory.---}Having established a phenomenological understanding of the enhancement of the SDE from the back action mechanism, we next turn our attention to a systematic self-consistent formalism. We switch to an action formalism and redefine $c_{\vec{k},\eta}$ in \equref{CouplingToVPandSC} as Grassmann variables depending on imaginary time $\tau$. We start from the effective action
\begin{equation}
    \mathcal{S} = \int \diff \tau \Biggl[ \sum_{\vec{k},\eta} c^\dagger_{\vec{k},\eta} \partial_\tau c^\pdagger_{\vec{k},\eta} + \mathcal{H}_c + \mathcal{S}_{\Delta} + \mathcal{S}_{\Phi} \Biggr], \label{EffectiveAction}
\end{equation}
that captures the desired phenomenology in a minimal setting. In \equref{EffectiveAction}, $\mathcal{S}_{\Delta}= \sum_{\vec{q}}(\frac{1}{g_{c}}|\Delta_{\vec{q}}|^{2}  + u_{\Delta} |\Delta_{\vec{q}}|^{4})$ and $\mathcal{S}_{\Phi} = \frac{1}{g_{v}}\Phi_{V}^{2}  + v_{\Phi_{V}}\Phi_{V}^4$ are the bare actions of the Hubbard-Stratonovich fields associated with superconductivity and valley polarization, respectively. We include terms up to quartic order (which, e.g., arise after having integrated out electronic degrees of freedom at higher energies) to stabilize coexistence of these two orders, as is seen in experiment \cite{lin_zero-field_2022,zhang_valley_2023,scammell_theory_2022}.

The saddle-point equations for $\Phi_V$ and $\Delta_{\vec{q}}$ read as
\begin{subequations}\begin{equation}
  \frac{\Phi_{V}}{g_{v}} + 2v_{\Phi} \Phi_{V}^{3} = \frac{1}{4}\sum_{\vec{k},p=\pm} p \tanh{\frac{ \mathcal{E}_{\vec{k},\vec{q},p}}{2T}},
  \label{eq:saddlephi}
\end{equation} 
\begin{equation}
    \frac{1}{g_{c}}  + 2 u_{\Delta}|\Delta_{\mathbf{q}}|^{2} = \sum_{\vec{k}} \frac{(\tanh{\frac{\mathcal{E}_{\vec{k},\vec{q},+}}{2 T}}+\tanh{\frac{\mathcal{E}_{\vec{k},\vec{q},-}}{2 T}})}{2(\mathcal{E}_{\vec{k},\vec{q},+}+\mathcal{E}_{\vec{k},\vec{q},-})},
    \label{eq:saddledel}
\end{equation}\label{SaddlePointEqs}\end{subequations}
or $\Delta_{\vec{q}}=0$, where $\mathcal{E}_{\vec{k},\vec{q},p} = \sqrt{\zeta_{\vec{k},\vec{q},+}^2 + |\Delta_{\vec{q}}|^2} + p \, \zeta_{\vec{k},\vec{q},-}$ with $\zeta_{\vec{k},\vec{q},\pm} =  (E_{\vec{k},\vec{q},+} \pm E_{\vec{k},\vec{q},-})/2$. As required, \equref{eq:saddledel} becomes equivalent to $a^{\text{S}}_{\vec{q}}=0$ when expanded to leading order in $\Delta_{\vec{q}}$.

\begin{figure}[tb]
   \centering
    \includegraphics[width=\linewidth]{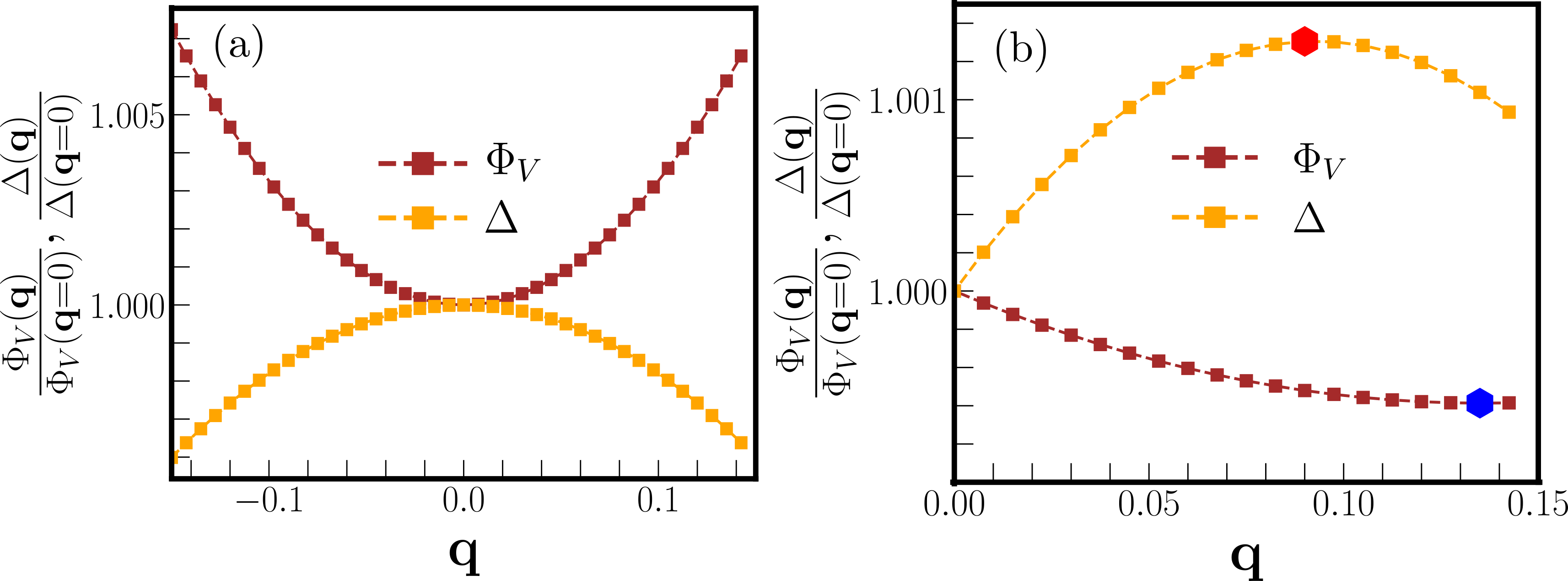}
        \caption{Coupling of supercurrent to valley polarization within microscopic theory. $\Phi_V(\vec{q})$ and $\Delta(\vec{q})$ along one-dimensional momentum cut with angle $\pi/6$ relative to the $q_x$ axis (a) without form factors ($f_{\vec{k}}=1$) and (b) with form factors, $f_{\vec{k}} =\sum_{j=1}^{3} \sin{\vec{a}_{j}\cdot\vec{k}}$. The red (blue) hexagon denotes the maximum (minimum) of $\Delta({\vec{q}})$ and $\Phi_{V}(\vec{q})$, respectively. We use $\beta=0.6$, $\mu = -1.36$, $g_{c}=5.6$, $g_{v}= 11$, $u_{\Delta}=0.25$, $v_{\Phi}=0.125$, $T=0.4$.}
    \label{fig:CurrentCouplingInMicroscopicTheory}
\end{figure} 

We solve \equref{SaddlePointEqs} self-consistently for both $\Delta_{\vec{q}}$ and $\Phi_V$ for given $\vec{q}$. This captures the mutual influence of these two orders, as $\Delta_{\vec{q}}$ and $\Phi_V$ enter each other's saddle-point equation, \equsref{eq:saddlephi}{eq:saddledel}, respectively, via the dispersion $\mathcal{E}$. Broken $C_{3z}$ symmetry is entirely encoded in the parameter $\beta$ entering the bare normal-state dispersion $\xi_{\vec{k}}$. Generically, the two coupled saddle-point equations are expected to capture any symmetry-allowed coupling term between $\Delta_{\vec{q}}$ and $\Phi_V$ in an effective free-energy expansion, including those in \equref{freenergyexpansion} but also terms involving higher powers of the order parameters. It turns out, though, that for the simple coupling in $\mathcal{H}_c$, where valley polarization only enters as an imbalance of the chemical potential in the two valleys, we find $\alpha=0$ in \equref{CouplingFunction} for any $\beta$. This is readily shown by expanding \equref{eq:saddlephi} in terms of $\delta \Phi_{V}(\vec{q}) = \Phi_{V}(\vec{q}) - \Phi_{V}(0)$ at fixed $\Delta(\vec{q})$, see \appref{CouplingValleyCurr}. This is why the back action is dominated by  the term $\alpha''|\Delta_{\vec{q}}|^2 \Phi_V^2$ in \equref{CouplingFunction} for small $\vec{q}$ as can be seen in \figref{fig:CurrentCouplingInMicroscopicTheory}(a), where $\delta \Phi_{V}(\vec{q}) \propto \delta \Delta(\vec{q}) \sim \vec{q}^2$ for small $\vec{q}$. We have checked that generalizing the coupling to $\Phi_V \sum_{\vec{k},\eta} \eta f_{\vec{k}\cdot\eta} c^\dagger_{\vec{k},\eta}c^\pdagger_{\vec{k}, \eta}$, with $C_{3z}$-invariant form factors $f_{\vec{k}}$, leads to finite $\alpha$; this is clearly visible in \figref{fig:CurrentCouplingInMicroscopicTheory}(b), where $\Phi_V(\vec{q})$ is found to have a finite slope at the maximum (red hexagon) of $\Delta(\vec{q})$.

\begin{figure}[tb]
   \centering
    \includegraphics[width=\linewidth]{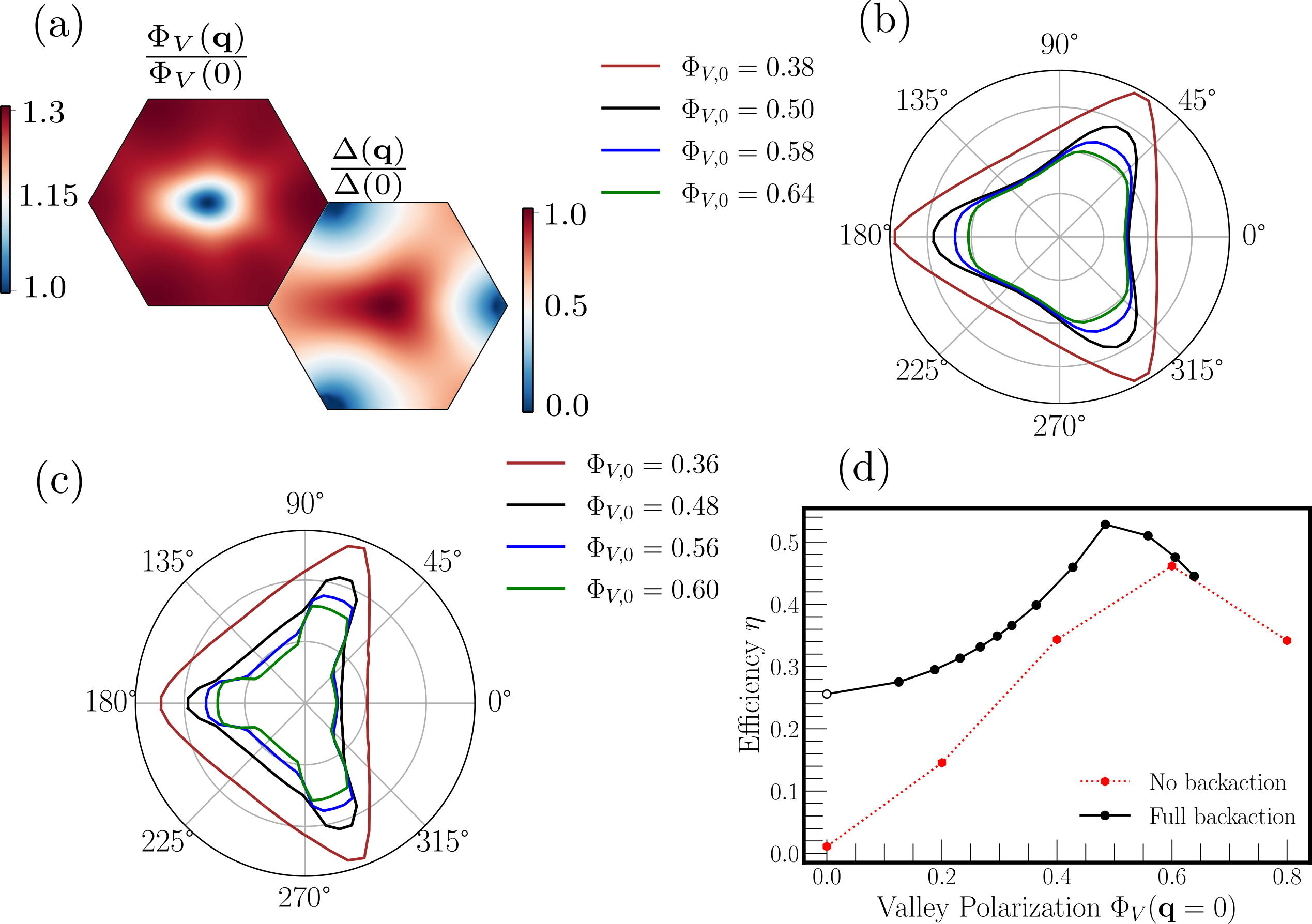}
    \caption{SDE within self-consistent theory, \equref{EffectiveAction}. (a) shows $\Phi_{V}(\vec{q})$ and $\Delta(\vec{q})$ in the upper and lower panel, for $|\Phi_{V}(\vec{q}=0)|=0.48$ and $\beta=0.6$. Resulting angular dependence of $J_c$ is shown without strain in (b) and with strain ($\beta=0.6$) in (c). (d) Efficiency $\eta$ as a function of $\Phi_V(\vec{q}=0)$ in presence of strain with back action (black solid line) and without back action (red dotted line). The parameters for $g_{c}$, $T$, $\mu$, $v_{\Phi}$, $u_{\Delta}$ are the same as in \figref{fig:CurrentCouplingInMicroscopicTheory}; only $g_{v}$ is varied to change $\Phi_V(\vec{q}=0)$.}
    \label{fig:SDEInMicroscopicTheory}
\end{figure} 

\vspace{1em}
\textit{SDE in self-consistent theory.---}To demonstrate that the enhancement mechanism for the SDE based on the back action is a robust phenomenon, we here discuss it for the original from of the order parameter $\Phi_V$ without form factors, $f_{\vec{k}}=1$, where the leading coupling ($\propto \alpha$) in the phenomenological theory in \equref{freenergyexpansion} is absent. In \figref{fig:SDEInMicroscopicTheory}(a), we plot $\Phi_V(\vec{q})$ and $\Delta(\vec{q})$, obtained by solving \equref{SaddlePointEqs}. We can clearly see the supercurrent-induced change of valley polarization, which is manifestly not $C_{3z}$ invariant, due to $\beta \neq 0$, and exhibits the strongest asymmetry under $\vec{q}\rightarrow -\vec{q}$ for $\vec{q}$ along $q_x$. Due to the additional coupling terms beyond those in \equref{freenergyexpansion}, $\Phi_V(\vec{q})$ looks different at large $\vec{q}$ from that shown in \figref{fig:PhenTheoryCubicCoupling}(a). However, the for our purpose more important $\Delta(\vec{q})$ looks very similar to $\Gamma_{\vec{q}}$ in \figref{fig:PhenTheoryCubicCoupling}(b). Evaluating the thermal expectation value of the current operator within the full theory (see \appref{microscopiccurrent}) as a function of $\vec{q}$, we compute the directional dependence of the critical current, shown in \figref{fig:SDEInMicroscopicTheory}(b) and (c) for $\beta=0$ and $\beta\neq 0$, respectively. Exactly as in the phenonemological theory, we find that broken $C_{3z}$ symmetry can enhance the SDE and that the current asymmetry first increases with small $\Phi_V$ (as significant TRS breaking is needed to generate a large current asymmetry) but then eventually decreases with large $\Phi_V$ (as it destabilizes superconductivity and also weakens the back action). This is also visible in \figref{fig:SDEInMicroscopicTheory}(d), where we plot the SDE efficiency $\eta$ of the full self-consistent theory (black line) as a function of valley polarization at $\vec{q}=0$ and compare it with the situation without back action (red dashed line); for the latter, we just fix $\Phi_V$ at the indicated value $\Phi_V(\vec{q}=0)$, independent of $\vec{q}$ and solve for superconductivity via \equref{ParticleParticleBubble}. We observe that the back action not only enhances, as before, the maximum value of $\eta$ that can be achieved, but also increases the efficiency significantly at smaller values of valley polarization. The reason for this amplification is the aforementioned dominance of the $\alpha''$-like couplings in the self-consistent theory, which enhance $\Phi_V(\vec{q})$ when suppressing $\Delta_{\vec{q}}$ with finite $\vec{q}$, see also \figref{fig:CurrentCouplingInMicroscopicTheory}(a). So even if we have very small $\Phi_V(\vec{q}=0)$, valley polarization can reach sizable values at non-zero $\vec{q}$ due to the coupling to the supercurrent, which in turn enhances the diode effect. This is why $\eta$ reaches finite values in the limit $\Phi_V(\vec{q}=0)\rightarrow 0^+$ with back action, while it has to vanish linearly with $\Phi_V(\vec{q}=0)$ without it in \figref{fig:SDEInMicroscopicTheory}(d). It shows that the proposed back action mechanism also enhances the \textit{typical} $\eta$ which can reach large values without the need of fine-tuning $\Phi_V(\vec{q}=0)$ to its optimal strength.

\vspace{1em}
\textit{Conclusion.---}We have shown, both using a simple free-energy expansion and a self-consistent theory, that the supercurrent can couple to and, hence, affect the valley polarization, $\Phi_V$, which in turn can enhance the SDE efficiency significantly, see \figref{fig:PhenTheoryLinearCoupling}(d) and \figref{fig:SDEInMicroscopicTheory}(d). Motivated by \refcite{lin_zero-field_2022}, this was formulated for a theory applicable to graphene-based systems; however, we expect that the basic mechanism based on coupling of the supercurrent to a TRS-breaking order parameter is more generally valid for zero-field superconducting diodes.

\begin{acknowledgments}
S.B.~and M.S.S.~acknowledge funding by the European Union (ERC-2021-STG, Project 101040651---SuperCorr). Views and opinions expressed are however those of the authors only and do not necessarily reflect those of the European Union or the European Research Council Executive Agency. Neither the European Union nor the granting authority can be held responsible for them. M.S.~thanks J.I.A.~Li and H.~Scammell for insightful discussions and comments on the manuscript. S.B. is grateful for discussions with J.A.~Sobral, A.~Rastogi, B.~Putzer and P.~Wilhelm.
\end{acknowledgments}

\bibliography{SCDiodeEffect}

\onecolumngrid

\begin{appendix}

\section{Details of models}\label{AppPhenTheory}
\begin{figure}[H]
   \centering
    \includegraphics[width=0.5\linewidth]{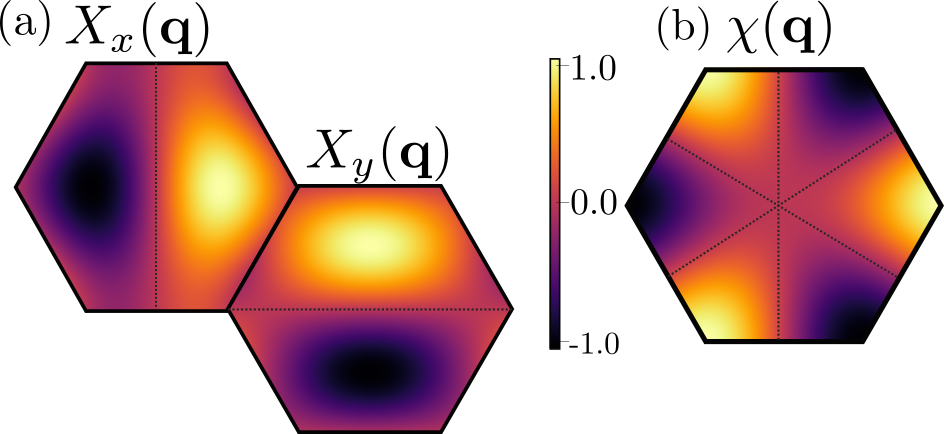}
    \caption{Momentum dependence of the two components of the function (a) $\vec{X}(\vec{q})$ and of (b) $\chi(\vec{q})$ as defined by \equref{Basisfunctions}. The black dashed lines indicate the lines across which the function changes sign.}
    \label{fig:basis1a}
\end{figure}

We here state the explicit form of the basis functions used to parameterize the phenomenological coupling in \equref{CouplingFunction} of the main text. Applying the convention, where one of the primitive lattice vectors of the underlying triangular lattice is given by $\vec{a}_1 = (1,0)^T$, these functions read as
\begin{align}
    \vec{X}(\vec{q}) &=  \left(\frac{2}{3}(2\cos{\frac{q_{x}}{2}} + \cos{\frac{\sqrt{3}q_{y}}{2}})\sin{\frac{q_{x}}{2}} \right., \left.{}\frac{2}{\sqrt{3}}\cos{\frac{q_{x}}{2}}\sin{\frac{\sqrt{3}q_{y}}{2}}\right), \\
   \chi(\vec{q}) &= 16 \left(-\cos{\frac{q_{x}}{2}} + \cos{\frac{\sqrt{3}q_{y}}{2}}\right)\sin{\frac{q_{x}}{2}}.  
   \label{Basisfunctions}
\end{align}
These are the leading (in the sense of having the fewest number of nodes/nodal lines) functions which obey the required symmetry constraints while being periodic on the Brillouin zone. They are illustrated in \figref{fig:basis1a}.

\section{Derivation of the saddle-point equations}\label{AppSaddleTheory}
To derive saddle point equations, we generalize \equref{EffectiveAction} to include non-trivial form factors, $f_{\vec{k}} \neq \text{const.}$, for the coupling of valley polarization to the electrons. The associated action in Matsubara-frequency notation becomes
\begin{equation}
    \mathcal{S}[\Delta_{\vec{q}},\Phi_{V}] = \frac{1}{g_{v}}\Phi_{V}^{2}+\frac{1}{g_{c}}|\Delta_{\vec{q}}|^{2}+ v_{\Phi_{V}}\Phi_{V}^4 
    + u_{\Delta} |\Delta_{\vec{q}}|^{4} + \sum_{k}\Psi_{\vec{k,q}}^{\dagger} \mathcal{G}^{-1}_{\vec{k,q}}\Psi_{\vec{k,q}},
\end{equation}
where the sum is done over $k=(\vec{k},\omega_n)$, comprising momentum $\vec{k}$ and Matsubara frequencies $\omega_n$.  $\mathcal{G}_{k,\vec{q}}$ and $\Psi_{k,\vec{q}}$ are given by 
\begin{equation}
    \mathcal{G}^{-1}_{k,\vec{q}} = \begin{bmatrix}
        -i\omega_{n} + \xi_{\vec{k}+\frac{\vec{q}}{2},+}+f_{\vec{k}+\frac{\vec{q}}{2},+}\Phi_{V} & \Delta_{\vec{q}}\\ 
         \Delta_{\vec{q}}^{\dagger}& -i\omega_{n} - \xi_{-\vec{k}+\frac{\vec{q}}{2},-}+f_{-\vec{k}+\frac{\vec{q}}{2},-}\Phi_{V}
        \end{bmatrix},\quad \Psi_{k,\vec{q}} = \binom{c_{\mathbf{k+\frac{q}{2},+}}}{c_{\mathbf{-k+\frac{q}{2},-}}^{\dagger}}. \label{DefinitionOfGinv}
\end{equation}
Integrating out the fermions, we obtain the effective action as, 
\begin{equation}
    \mathcal{S}_{\text{eff}} = \frac{\beta}{g_{v}}\Phi_{V}^{2}+\frac{\beta}{g_{c}}|\Delta_{\vec{q}}|^{2}+ \beta v_{\Phi_{V}}\Phi_{V}^4 
    + \beta u_{\Delta} |\Delta_{\vec{q}}|^{4} - \sum_{k}\text{tr}\ln{\mathcal{G}^{-1}_{\vec{k,q}}}.
\end{equation}
Now, we calculate the saddle point equation for $\Delta_{\vec{q}}$ we set $  \delta S_{\text{eff}}/\delta \Delta_{\vec{q}} \overset{!}{=} 0$
\begin{align}
    \beta \frac{\Delta_{\vec{q}}^{\dagger}}{g_{c}} + 2 \beta u_{\Delta}|\Delta_{\vec{q}}|^{2}\Delta_{\vec{q}}^{\dagger} -  \sum_{k}\text{tr}\left[\mathcal{G}_{\vec{k,q}} \frac{\delta}{\delta \Delta_{\vec{q}}} \begin{pmatrix}
       -i\omega_{n}+E_{\vec{k,q},+} & \Delta_{\vec{q}}\\
          \Delta_{\vec{q}}^{\dagger}& -i\omega_{n}-E_{\vec{k,q},-}
         \end{pmatrix}\right] &= 0 \\
         \beta \frac{\Delta_{\vec{q}}^{\dagger}}{g_{c}} + 2 \beta u_{\Delta}|\Delta_{\vec{q}}|^{2}\Delta_{\vec{q}}^{\dagger} -  \sum_{k}\text{tr}\left[\mathcal{G}_{\vec{k,q}} \begin{pmatrix}
            0 & 1\\
               0& 0
              \end{pmatrix}\right] &= 0 
\end{align}
Now,
\begin{equation}
    \mathcal{G}_{\vec{k,q}} = \frac{-1}{(-i\omega_{n} + E_{\vec{k,q}, +})(-i\omega_{n} - E_{\vec{k,q}, -})-\Delta_{\vec{q}}^{2}}\begin{bmatrix}
        i\omega_{n} + E_{\vec{k,q},-} & \Delta_{\vec{q}}\\ 
         \Delta_{\vec{q}}^{\dagger}& i\omega_{n} - E_{\vec{k,q},+}
        \end{bmatrix}
\end{equation}
where we have redefined our energies as, 
\begin{equation}
    E_{\vec{k,q}, +} = \xi_{\vec{k}+\frac{\vec{q}}{2},+} + f_{\vec{k} + \frac{\vec{q}}{2},+} \Phi_{V}, \quad   E_{\vec{k,q}, -} = \xi_{-\vec{k}+\frac{\vec{q}}{2},-} - f_{-\vec{k} + \frac{\vec{q}}{2},-} \Phi_{V}, 
\end{equation}
which is the generalization of the energies defined below \equref{ParticleParticleBubble} of the main text to include form factors. 
Therefore, the trace term reduces to
\begin{equation}
    \sum_{k} \text{tr}\left[\mathcal{G}_{\vec{k,q}}  \begin{pmatrix}
        0 & 1\\
           0& 0
          \end{pmatrix}\right] =\sum_{\vec{k},n} \frac{\Delta_{\vec{q}}^{\dagger}}{\Delta_{\vec{q}}^{2} + \zeta_{\vec{k,q},+}^{2}+ (\omega_{n} + i\zeta_{\vec{k,q},-})^{2}}, \quad \text{with} \quad \zeta_{\vec{k},\vec{q},\pm} =  (E_{\vec{k},\vec{q},+} \pm E_{\vec{k},\vec{q},-})/2.
\end{equation}
Performing the sum over Matsubara frequencies, we get
\begin{equation}
    \text{tr}\left[\mathcal{G}_{\vec{k,q}}  \begin{pmatrix}
        0 & 1\\
           0& 0
          \end{pmatrix}\right] = \sum_{\vec{k}} \frac{\Delta_{\vec{q}}^{\dagger}(\tanh{\frac{\mathcal{E}_{\vec{k},\vec{q},+}}{2 T}}+\tanh{\frac{\mathcal{E}_{\vec{k},\vec{q},-}}{2 T}})}{2T(\mathcal{E}_{\vec{k},\vec{q},+}+\mathcal{E}_{\vec{k},\vec{q},-})}
\end{equation}
where $\mathcal{E}_{\vec{k,q},p} = \sqrt{\zeta_{\vec{k},\vec{q},+}^2 + |\Delta_{\vec{q}}|^2} + p \, \zeta_{\vec{k},\vec{q},-}$. 
Consequentially, this leads us to our saddle point equation for $\Delta_{\vec{q}}$,
\begin{equation}
    \frac{\Delta_{\vec{q}}^{\dagger}}{g_{c}}  + 2 u_{\Delta}|\Delta_{\mathbf{q}}|^{2}\Delta_{\vec{q}}^{\dagger} = \sum_{\vec{k}} \frac{\Delta_{\vec{q}}^{\dagger}(\tanh{\frac{\mathcal{E}_{\vec{k},\vec{q},+}}{2 T}}+\tanh{\frac{\mathcal{E}_{\vec{k},\vec{q},-}}{2 T}})}{2(\mathcal{E}_{\vec{k},\vec{q},+}+\mathcal{E}_{\vec{k},\vec{q},-})}
\end{equation}  
as stated in \equref{eq:saddledel} of the main text.
Moving onto the saddle point equation for valley polarization, we start similarly by setting $  \delta S_{\text{eff}}/\delta \Phi_{V} \overset{!}{=} 0$ 
\begin{align}
    \frac{2\beta\Phi_{V}}{g_{v}} + 4v_{\Phi}\beta\Phi_{V}^{3} -   \sum_{k}\text{tr}\left[\mathcal{G}_{\vec{k,q}} \frac{\delta}{\delta \Phi_{V}} \begin{pmatrix}
        -i\omega_{n}+E_{\vec{k,q},+} & \Delta_{\vec{q}}\\
           \Delta_{\vec{q}}^{\dagger}& -i\omega_{n}-E_{\vec{k,q},-} 
          \end{pmatrix}\right] &= 0 \\
          \frac{2\beta\Phi_{V}}{g_{v}} + 4v_{\Phi}\Phi_{V}^{3} -   \sum_{k}\text{tr}\left[\mathcal{G}_{\vec{k,q}}\begin{pmatrix}
            \mathcal{F}_{\vec{k,q},+} & 0\\
              0& \mathcal{F}_{\vec{k,q},-} 
              \end{pmatrix}\right] &= 0
\end{align}
where $\mathcal{F}_{\vec{k},\vec{q},+ } =  f_{\vec{k} + \frac{\vec{q}}{2},+}$ and $\mathcal{F}_{\vec{k},\vec{q},- } =  f_{-\vec{k} + \frac{\vec{q}}{2},-}$. 
Evaluating the trace, we get
\begin{equation}
    \sum_{k}\text{tr}\left[\mathcal{G}_{\vec{k,q}}\begin{pmatrix}
        \mathcal{F}_{\vec{k,q},+} &0\\
      0& \mathcal{F}_{\vec{k,q},-}\end{pmatrix}\right] = \sum_{\vec{k},n} \frac{(i\omega_{n}+E_{\vec{k,q},-}) \mathcal{F}_{\vec{k,q},+} +(i\omega_{n}-E_{\vec{k,q},+}) \mathcal{F}_{\vec{k,q},-}}{\Delta_{\vec{q}}^{2} + 
           \zeta_{\vec{k,q},+}^{2}+ (\omega_{n} + i\zeta_{\vec{k,q},-})^{2}}
    \end{equation}
If we now evaluate the Matsubara sums, we arrive at
\begin{equation}
    \sum_{k}\text{tr}\left[\mathcal{G}_{\vec{k,q}}\begin{pmatrix}
        \mathcal{F}_{\vec{k,q},+} & 0\\
         0& \mathcal{F}_{\vec{k,q},-}\end{pmatrix}\right] =\frac{1}{2} \sum_{\vec{k}} \left(F_{\vec{k,q},+} \tanh{\frac{\mathcal{E}_{\vec{k},\vec{q},+}}{2 T}}-F_{\vec{k,q},-}\tanh{\frac{\mathcal{E}_{\vec{k},\vec{q},-}}{2 T}}\right) 
\end{equation}
where the coefficients are given by 
\begin{equation}
    F_{\vec{k,q},+} = \mathcal{F}_{\vec{k,q}+}l_{\vec{k,q}} + \mathcal{F}_{\vec{k,q}-}m_{\vec{k,q}}, \quad F_{\vec{k,q},-} = \mathcal{F}_{\vec{k,q}-}l_{\vec{k,q}} + \mathcal{F}_{\vec{k,q}+}m_{\vec{k,q}} 
\end{equation}
\begin{equation}
    l_{\vec{k,q}} = \frac{1}{2}\left(1+\frac{\zeta_{\vec{k,q},+}}{\sqrt{\zeta_{\vec{k,q},+}^{2}+\Delta_{\vec{q}}^{2}}}\right), \quad  m_{\vec{k,q}} = \frac{1}{2}\left(1-\frac{\zeta_{\vec{k,q},+}}{\sqrt{\zeta_{\vec{k,q},+}^{2}+\Delta_{\vec{q}}^{2}}}\right) 
\end{equation}
Therefore, we obtain our saddle point equation for valley polarization as, 
\begin{equation}
    \frac{\Phi_{V}}{g_{v}} + 2v_{\Phi}\Phi_{V}^{3} = \frac{1}{4}\sum_{\vec{k}} \left(F_{\vec{k,q},+} \tanh{\frac{\mathcal{E}_{\vec{k},\vec{q},+}}{2 T}}-F_{\vec{k,q},-}\tanh{\frac{\mathcal{E}_{\vec{k},\vec{q},-}}{2 T}}\right) 
\end{equation} 
which reduces to \equref{eq:saddlephi} when the form factors are trivial [$f_{\vec{k},\pm}=1$ in \equref{DefinitionOfGinv}]. 

\section{Coupling of supercurrent and valley polarization within full self-consistent theory}\label{CouplingValleyCurr}
In this part of the appendix, we elucidate the impact of finite $\vec{q}$ on superconductivity and valley polarization within the coupled non-linear self-consistency equations in \equref{SaddlePointEqs} of the main text. We choose a gauge where $\Delta(\vec{q})\in\mathbb{R}$, for notational simplicity, and begin with \equref{eq:saddlephi} which we expand to linear order in $\vec{q}$, $\delta\Phi_V(\vec{q}):= \Phi_V(\vec{q})-\Phi_V(\vec{q}=0)$, and $\delta\Delta(\vec{q}):= \Delta(\vec{q})-\Delta(\vec{q}=0)$. This leads to
\begin{equation}
    C_{\Phi \Phi}  \delta\Phi_V(\vec{q}) + C_{\Phi \Delta}  \delta\Delta(\vec{q}) = \vec{n}_\Phi \cdot \vec{q} + \mathcal{O}(\vec{q}^2,\delta\Phi_V^2,\delta\Delta^2), \label{ExpandEqForPhi}
\end{equation}
where 
\begin{align}
    C_{\Phi \Phi} &= \frac{1}{g_v} + 6v_\Phi \Phi_V^2(\vec{q}=0) - \frac{1}{8T} \sum_{\vec{k},p} \frac{1}{\cosh^2(\frac{\mathcal{E}_{\vec{k},\vec{q}=0,p}}{2T})}, \\
    C_{\Phi \Delta} &= -\frac{\Delta(\vec{q}=0)}{8T} \sum_{\vec{k},p}\frac{1}{\sqrt{\xi_{\vec{k}}^2+\Delta^2(\vec{q}=0)}} \frac{p}{\cosh^2(\frac{\mathcal{E}_{\vec{k},\vec{q}=0,p}}{2T})}, \\
    \vec{n}_{\Phi} &= \frac{1}{16 T} \sum_{\vec{k},p} \frac{\vec{\nabla}_{\vec{k}}\xi_{\vec{k}}}{\cosh^2(\frac{\mathcal{E}_{\vec{k},\vec{q}=0,p}}{2T})}.
\end{align}
In our case, $\mathcal{E}_{\vec{k},\vec{q}=0,p} = \sqrt{\xi_{\vec{k}}^2 + \Delta^2(\vec{q}=0)} + p \Phi_{V}(\vec{q}=0)$ only depends on $\vec{k}$ through $\xi_{\vec{k}}$, leading to $\vec{n}_{\Phi}=0$. Note this would generically be different if the order parameter of superconductivity or valley polarization [$f_{\vec{k}} \neq \text{const.}$ in \equref{DefinitionOfGinv}] depended on momentum. We thus conclude that $\delta\Phi_V(\vec{q}) \propto \delta\Delta(\vec{q}) + \mathcal{O}(\vec{q}^2)$. This is consistent with \figref{fig:CurrentCouplingInMicroscopicTheory}(a) and $\alpha=0$ in \equref{CouplingFunction} of the main text.  

We can learn more by expanding \equref{eq:saddledel} in the same way,
\begin{equation}
    C_{\Delta \Delta}  \delta\Delta(\vec{q}) + C_{\Delta \Phi}  \delta\Phi(\vec{q}) = \vec{n}_\Delta \cdot \vec{q} + \mathcal{O}(\vec{q}^2,\delta\Phi_V^2,\delta\Delta^2), \label{ExpandEqForDelta}
\end{equation}
where the coefficients are given by
\begin{align}
    C_{\Delta \Delta} &= 4u_\Delta \Delta(\vec{q}=0) + \frac{1}{2}  \sum_{\vec{k},p}\tanh\left(\frac{\mathcal{E}_{\vec{k},\vec{q}=0,p}}{2T}\right)  \frac{\Delta(\vec{q}=0)}{(\xi_{\vec{k}}^2 + \Delta^2(\vec{q}=0))^{\frac{3}{2}}}  - \frac{1}{4T} \sum_{\vec{k},p} \frac{\Delta(\vec{q}=0)}{\xi_{\vec{k}}^2 + \Delta^2(\vec{q}=0)} \frac{1}{\cosh^2(\frac{\mathcal{E}_{\vec{k},\vec{q}=0,p}}{2T})}, \\
    C_{\Delta\Phi } &= -\frac{1}{2T} \sum_{\vec{k},p}\frac{1}{\sqrt{\xi_{\vec{k}}^2+\Delta^2(\vec{q}=0)}} \frac{p}{\cosh^2(\frac{\mathcal{E}_{\vec{k},\vec{q}=0,p}}{2T})}, \\
    \vec{n}_{\Delta} &= \frac{1}{8 T} \sum_{\vec{k},p} \frac{1}{\sqrt{\xi_{\vec{k}}^2+\Delta^2(\vec{q}=0)}} \frac{p\vec{\nabla}_{\vec{k}}\xi_{\vec{k}}}{\cosh^2(\frac{\mathcal{E}_{\vec{k},\vec{q}=0,p}}{2T})}.
\end{align}
The same comments as those above for $\vec{n}_{\Phi}$ also apply for $\vec{n}_{\Delta}$, which thus vanishes, $\vec{n}_{\Delta}=0$.
Without fine-tuning, we expect $C_{\Phi\Phi} C_{\Delta \Delta} \neq C_{\Phi\Delta}C_{\Delta\Phi}$ and \equsref{ExpandEqForPhi}{ExpandEqForDelta} therefore imply $\delta \Delta(\vec{q}), \delta \Phi_V(\vec{q}) = \mathcal{O}(\vec{q}^2)$ exactly as we find numerically, see \figref{fig:CurrentCouplingInMicroscopicTheory}(a). This is the reason why the maximum of $\Delta(\vec{q})$ [minimum of $\Phi_V(\vec{q})$] can be pinned to $\vec{q}=0$ even when $C_{3z}$ is broken. As $\vec{n}_{\Phi}, \vec{n}_{\Delta}$ become non-zero when form factors are introduced, the maximum of $\Delta(\vec{q})$ [minimum of $\Phi_V(\vec{q})$] move to finite values when $C_{3z}$ is broken as is observed in \figref{fig:CurrentCouplingInMicroscopicTheory}(b).

\section{Microscopic calculation of current}\label{microscopiccurrent}
We can write a BdG Hamiltonian for our system as, 
\begin{equation}
    H_{\text{BdG}} = \begin{bmatrix}
    E_{\mathbf{k+\frac{q}{2}},+} &\Delta_{\mathbf{q}} \\ 
    \Delta_{\mathbf{q}}^{*} & -E_{\mathbf{-k+q/2},-} 
    \end{bmatrix},  \quad \Psi = \binom{c_{\mathbf{k+\frac{q}{2},+}}}{c_{\mathbf{-k+\frac{q}{2},-}}^{\dagger}}, \quad \Psi^{\dagger} = \begin{pmatrix}
     c_{\mathbf{k+\frac{q}{2},+}}^{\dagger}& c_{\mathbf{-k+\frac{q}{2},-}}
    \end{pmatrix}
    \end{equation}
and    
\begin{equation}
        \binom{c_{\mathbf{k+\frac{q}{2},+}}}{c_{\mathbf{-k+\frac{q}{2},-}}^{\dagger}}  = \begin{pmatrix}
        u_{\mathbf{k,q}} & v_{\mathbf{k,q}} \\ -v_{\mathbf{k,q}}^{*}
         & u_{\mathbf{k,q}}^{*}
        \end{pmatrix}\binom{\gamma_{\mathbf{k,q},+}}{\gamma^{\dagger}_{\mathbf{-k,q},-}}.
        \label{eq:bf_transformation}
        \end{equation}
The BdG Hamiltonian in diagonal basis can be now written as,
\begin{equation}
    H_{\text{BdG}} = UH_{\text{diag}}U^{\dagger}, \quad U = \begin{pmatrix}
        u_{\mathbf{k,q}} & v_{\mathbf{k,q}} \\ -v_{\mathbf{k,q}}^{*}
         & u_{\mathbf{k,q}}^{*}
        \end{pmatrix}, \quad U^{\dagger} = \begin{pmatrix}
            u_{\mathbf{k,q}}^{*} & -v_{\mathbf{k,q}} \\ v_{\mathbf{k,q}}^{*}
             & u_{\mathbf{k,q}}
            \end{pmatrix}  
\end{equation}

The expression for the current (see, e.g., \cite{daido_intrinsic_2022}) reads as
\begin{align}
    J &= \sum_{\mathbf{k}}\text{tr}\left[\partial_{\mathbf{q}}H_{\text{N}}n_{F}(H_{\text{BdG}})\right]\\
      &= \sum_{\mathbf{k}}\text{tr}\left[\begin{pmatrix} \partial_{\mathbf{q}}E_{\mathbf{k+\frac{q}{2}},+} &0 \\ 
        0& -\partial_{\mathbf{q}}E_{\mathbf{-k+q/2},-} 
        \end{pmatrix}n_{F}(H_{\text{BdG}})\right],
\end{align}
where
 $n_{F}(\epsilon) = (1+e^{\beta\epsilon})^{-1}$  is the Fermi distribution function and
 \begin{equation}
    H_{\text{N}} = \begin{pmatrix}
    E_{\mathbf{k+\frac{q}{2}},+} &0 \\ 
    0 & -E_{\mathbf{-k+q/2},-} 
    \end{pmatrix}.
\end{equation}
Therefore we have, 
\begin{align}
    n(H_{\text{BdG}}) &= Un_{F}(H_{\text{diag}})U^{\dagger}\\
                      &= U \begin{pmatrix}
                        \frac{1}{1+e^{\beta E_{1,\mathbf{k,q}}}} & 0\\
                        0 & \frac{1}{1+e^{-\beta E_{2,\mathbf{k,q}}}} 
                      \end{pmatrix}U^{\dagger}
\end{align}
Plugging this into the expression for current we get:
\begin{align}
    J &= \sum_{\mathbf{k}}\text{tr}\left[U^{\dagger} \begin{pmatrix} \partial_{\mathbf{q}}E_{\mathbf{k+\frac{q}{2}},+} &0 \\ 
        0& -\partial_{\mathbf{q}}E_{\mathbf{-k+q/2},-} 
        \end{pmatrix}U n_{F}(H_{\text{diag}})\right]\\
      &= \sum_{\mathbf{k}}\text{tr}\left[\begin{pmatrix}
        u_{\mathbf{k,q}}^{*} & -v_{\mathbf{k,q}}\\ 
        v_{\mathbf{k,q}}^{*} & u_{\mathbf{k,q}}
        \end{pmatrix} \begin{pmatrix} u_{\mathbf{k,q}} \partial_{\mathbf{q}}E_{\mathbf{k+\frac{q}{2}},+}&v_{\mathbf{k,q}}\partial_{\mathbf{q}}E_{\mathbf{k+\frac{q}{2}},+} \\ 
            v_{\mathbf{k,q}}^{*}\partial_{\mathbf{q}}E_{\mathbf{-k+\frac{q}{2}},-}& -u^{*}_{\mathbf{k,q}}\partial_{\mathbf{q}}E_{\mathbf{-k+q/2},-}
            \end{pmatrix}n_{F}(H_{\text{diag}})\right]\\
      &=\sum_{\mathbf{k}}\text{tr}\left[\begin{pmatrix} |u_{\mathbf{k,q}}|^{2} \partial_{\mathbf{q}}E_{\mathbf{k+\frac{q}{2}},+}-|v_{\mathbf{k,q}}|^{2}\partial_{\mathbf{q}}E_{\mathbf{-k+\frac{q}{2}},-} &u_{\mathbf{k,q}}^{*}v_{\mathbf{k,q}}\partial_{\mathbf{q}}E_{\mathbf{k+\frac{q}{2}},+}+u^{*}_{\mathbf{k,q}}v_{\mathbf{k,q}}\partial_{\mathbf{q}}E_{\mathbf{-k+q/2},-} \\ 
        v_{\mathbf{k,q}}^{*}u_{\mathbf{k,q}} \partial_{\mathbf{q}}E_{\mathbf{k+\frac{q}{2}},+}+u_{\mathbf{k,q}}v_{\mathbf{k,q}}^{*}\partial_{\mathbf{q}}E_{\mathbf{-k+\frac{q}{2}},-}&  |v_{\mathbf{k,q}}|^{2} \partial_{\mathbf{q}}E_{\mathbf{k+\frac{q}{2}},+}-|u_{\mathbf{k,q}}|^{2}\partial_{\mathbf{q}}E_{\mathbf{-k+\frac{q}{2}},-} 
        \end{pmatrix}n_{F}(H_{\text{diag}})\right]        
\end{align}
\begin{multline}
    J =  \sum_{\mathbf{k}} (|u_{\mathbf{k,q}}|^{2} \partial_{\mathbf{q}}E_{\mathbf{k+\frac{q}{2}},+}-|v_{\mathbf{k,q}}|^{2}\partial_{\mathbf{q}}E_{\mathbf{-k+\frac{q}{2}},-})n_{F}(E_{1,\mathbf{k,q}})\\
    +\sum_{\mathbf{k}} (|v_{\mathbf{k,q}}|^{2} \partial_{\mathbf{q}}E_{\mathbf{k+\frac{q}{2}},+}-|u_{\mathbf{k,q}}|^{2}\partial_{\mathbf{q}}E_{\mathbf{-k+\frac{q}{2}},-})n_{F}(-E_{2,\mathbf{k,q}}) 
\end{multline}

\end{appendix}
\end{document}